\documentclass[3p,12pt]{elsarticle}

\usepackage{url}
\usepackage{amsbsy}
\usepackage{amssymb}
\usepackage{amsmath}
\usepackage[english]{babel}
\usepackage[numbers]{natbib}
\usepackage{comment}

\renewcommand*{\today}{7 August 2017}
\begin{document}

\begin{frontmatter}

\title{{Constructing an Explicit AdS/CFT Correspondence with Cartan Geometry}}
\let\today\relax
%\date{7 August 2017}

\author{Jeffrey S. Hazboun}
\ead{hazboun@uw.edu}
\fntext[fn1]{Received 7 August 2017, Revised 11 February 2018, Accepted 14 February 2018, Available online 19 February 2018}

\address{Center for Advanced Radio Astronomy, University of Texas Rio Grande Valley}

\address{Physical Sciences Division, University of Washington Bothell}

\begin{abstract}
 An explicit AdS/CFT correspondence is shown for the Lie group $SO(4,2)$. The Lie symmetry structures allow for the construction of two physical theories through the tools of Cartan geometry. One is a gravitational theory that has anti-de~Sitter symmetry. The other is also a gravitational theory but is conformally symmetric and lives on 8-dimensional biconformal space. These ``extra'' four dimensions have the degrees of freedom used to construct a Yang-Mills theory. The two theories, based on AdS or conformal symmetry, have a natural correspondence in the context of their Lie algebras alone where neither SUSY, nor holography, is necessary.
\end{abstract}


\end{frontmatter}

\section{Introduction}

The original AdS/CFT conjecture \cite{Maldacena:1997re}, now generalized and referred to as the gauge/gravity correspondence, has been widely successful at making connections with the realms of condensed matter physics \cite{Hartnoll:2008kx} and quark-gluon plasmas \cite{Erlich:2005qh}.
Here we show, starting with the Lie group (algebra) structure, how
to construct two theories (manifold, fields, metric, action principle)
with an exact correspondence that relates fields in one theory to those
in the other. 
The correspondence is given by the change of basis from AdS generators to conformal generators and constructed using Cartan geometry, which gives us the toolkit to build a manifold,
together with connections, curvatures and metric structure, from the
Lie symmetry structure. The correspondence will be between two theories
of external symmetries, i.e. gravitational theories, but links will be made with Yang-Mills theories\footnote{We define a Yang-Mills theory as having symmetry generators with a Euclidean inner product, distinct from the interesting use in \cite{Attard:2015jfr}, where the definition is more general.} via other well known results, both from the perspective of graviweak theories, \cite{Nesti:2007ka}, and from Euclidean general relativity \cite{Eguchi:1979yx}.
In the latter case the Yang-Mills theory depends on the appearance of a Euclidean inner product within the conformal theory, derived
from the Killing metric of the Lie algebra; a result known from previous
work in the gravitational gauge theory that gives biconformal space
\cite{Wheeler:1997pc}.  

A Cartan geometry uses the principal fiber bundle, constructed by
taking the quotient of a Lie group, $\mathcal{G}$, by one of its
Lie subgroups, $\mathcal{H}$, as a model space for a curved manifold. The Maurer-Cartan equation %,
%$\mathbf{d}\boldsymbol{\omega}^{A}=-\frac{1}{2}c_{\;BC}^{A}\boldsymbol{\omega}^{B}\wedge\boldsymbol{\omega}^{C}$,
gives a set of differential relations between the connections defined
via an association to the Lie algebra generators. The Cartan connection,
taking values in the Lie algebra, $\mathfrak{g}$, is a curved version of those connections \cite{Sharpe:1997a}. From a gravitational perspective
this procedure gives a generalization of Riemannian geometries where
the tangent spaces of the manifold are replaced by the homogeneous
space, constructed via the quotient $\mathcal{G}/\mathcal{H}$. This
construction goes back to Cartan \cite{Cartan1910a} and has been
used in the context of gravitational gauge theories since around 1960
\cite{Utiyama:1956p1235,Kibble:1961p1468}. Investigations into SUSY
gravitational theories renewed interest in these techniques in the
late 70s through the 80s \cite{MacDowell:1977p1566,Neeman:1978p1517,Neeman:1978p1521,Ivanov:1982p1172,Ivanov:1982p1201}.
For a modern construction see \cite{Sharpe:1997a,Hazboun:2013lra,Westman:2012zk,Wise:2006sm}.

Since the connections introduced in these geometries are equivalent to gauge fields, this
construction works equally well for constructing theories with internal
symmetries, i.e. particle theories. In the case of internal symmetries
the subgroup, $\mathcal{H}$, is usually thought of as a normal subgroup\footnote{An interesting result of the work described in this manuscript and explicitly derived by Lovelady and Wheeler \cite{Lovelady:2015xhh} is that a particle theory can be shown to originate from a non-normal subgroup.}, $\mathcal{G}=\mathcal{T}\times\mathcal{H}$. Cartan geometries can be used to write down both general relativity and the standard model of particle physics (or rather its non-quantized predecessor). 

In this manuscript we outline how to explicitly map the fields in
the AdS symmetric theory to those in the conformally symmetric theory
and present the technique for extending this calculation to other
symmetries. The mapping to a Yang-Mills theory relies on the choice of the conformal
gauging that gives biconformal space, described below. The original conjecture
is a correspondence 
between the degrees of freedom of an AdS spacetime and a conformally
super symmetric Yang-Mills theory, in the context of string
theory. The equivalence shown here differs in that it is
classical and non-SUSY, in contrast to the full blown gauge/gravity duality of
string theory, which includes quantization, supersymmetry (however,
see \cite{Anderson:2003db}), and holography. The methods used here may hint at a path towards a generic proof of the original conjecture.

\section{The Anti-de Sitter / Conformal Symmetry Correspondence}

In the present case we wish to construct two theories from the symmetries described by the Lie group
$SO(4,2)$, a 5-dimensional AdS-symmetric theory and a 4-dimensional
conformally symmetric theory. Since the original conjecture is posited in these dimensions, we use them where necessary for brevity. However the construction, and notation, are well suited for any $SO(n,2)$. The most straightforward
interpretation of $SO(4,2)$ is as the pseudo-rotation group in 6-dimensions,
known as the defining representation. The generators of any $SO\left(p,q\right)$
group can be written as $M_{\;B}^{A}$, where the indices denote the
plane of the pseudo-rotation and the upper index has been raised by
the orthonormal metric 
\[
\eta^{AB}\equiv diag(\underbrace{1,\dots,1}_{p}\underbrace{-1,\dots,-1}_{q}) .
\]
These generators are then viewed as the set of symmetries that keep
the form of $\eta^{AB}$ unchanged. In our present example we
use the metric in the following form, $\eta^{AB}\equiv diag\left(-1,1,1,1,1,-1\right)$.
The index $A$ runs from $0\dots5$, the typeface Latin indices from
the center of the alphabet, $\mathtt{i}$, run from $0\dots4$ and
the lower case Latin indices, $a$, run from $0\dots3$. The indices
are nested in the following manner, $A=\left(\mathtt{i},5\right)=\left(a,4,5\right)$. For clarity no indices will be raised or lowered; all metrics will be explicitly shown.

There is an accidental relationship between the pseudo-rotation generators
of 6-dimensions and those of both the AdS symmetries of a 5-dimensional
space and the conformal symmetries of a 4-dimensional space, which
all have the same Lie algebra, $\mathfrak{so}\left(4,2\right)$. Using
our nested notation for the indices we can write these generators
using three different bases of the Lie algebra, each representing
a different physical set of symmetries,

\begin{equation}\label{eq:LieAlgebraBases}
	\left.
	\begin{aligned}        
		M_{\;\mathtt{j}}^{\mathtt{i}}&\\        		\mathtt{P}_{\mathtt{i}}&\equiv\frac{1}{\ell}M_{\;\mathtt{i}}^{5}      
	\end{aligned} 
	\right\} 
	M_{\;B}^{A}
	\begin{cases}
		M_{\;b}^{a}\\
		P_{a}\equiv\frac{1}{\sqrt{2}}\left(M_{\;a}^{4}+\eta_{ab}M_{\;5}^{b}\right)\\
		K^{a}\equiv\frac{1}{\sqrt{2}}\left(M_{\;5}^{a}-\eta^{ab}M_{\;b}^{4}\right)\\
		D\equiv M_{\;5}^{4} .
	\end{cases} 
\end{equation} 
The generators in the center column give the set of pseudo-rotations
in a 6-dim space. Those on the left give the AdS symmetries of a 5-dim
space, where we have included the natural length scale, $\ell$, present
in an AdS geometry. Finally, those on the right are the natural basis
for the conformal symmetries of a 4-dim space. 

Every Lie group comes equipped with a $1$-form, called the Maurer-Cartan
(MC) form, $\boldsymbol{\theta}$, satisfying the equation, $\mathbf{d}\boldsymbol{\theta}=-\frac{1}{2}\left[\boldsymbol{\theta},\boldsymbol{\theta}\right]$.
This is usefully expanded in a Lie algebra basis $\mathbf{d}\boldsymbol{\omega}^{A}=-\frac{1}{2}c_{\;BC}^{A}\boldsymbol{\omega}^{B}\wedge\boldsymbol{\omega}^{C}$,
where $\boldsymbol{\theta}\equiv\boldsymbol{\omega}^{A}G_{A}$, $G_{A}$
are the Lie algebra generators and $c_{\;BC}^{A}$ are the structure
constants. The MC equation can be written in terms of any of the bases
written in Eqn (\ref{eq:LieAlgebraBases}), and are listed in Table
(\ref{MCEqnTable}) for reference.

\begin{table}[h]

\bgroup
\def\arraystretch{1.2}

\begin{center}
\hbox{
\hspace{0.8in}
\begin{tabular}{|c|}
\hline 
Pseudo-Rotations\tabularnewline
\hline 
$\boldsymbol{\theta}=\boldsymbol{\omega}_{\;B}^{A}M_{\;A}^{B}$\tabularnewline
\hline 
$\mathbf{d}\boldsymbol{\omega}_{\;B}^{A}=\boldsymbol{\omega}_{\;B}^{C}\wedge\boldsymbol{\omega}_{\;C}^{A}$\tabularnewline
\hline 
\end{tabular}%

\hspace{0.01in}
\begin{tabular}{|c|}
\hline 
Anti-de Sitter\tabularnewline
\hline 
$\boldsymbol{\theta}=\boldsymbol{\omega}_{\;\mathtt{i}}^{\mathtt{j}}M_{\;\mathtt{j}}^{\mathtt{i}}+\mathbf{E}^{\mathtt{i}}\mathtt{P}_{\mathtt{i}}$\tabularnewline
\hline 
$\mathbf{d}\boldsymbol{\omega}_{\;\mathtt{j}}^{\mathtt{i}}=\boldsymbol{\omega}_{\;\mathtt{j}}^{\mathtt{k}}\wedge\boldsymbol{\omega}_{\;\mathtt{k}}^{\mathtt{i}}+\frac{\epsilon}{\ell^{2}}\eta_{\mathtt{k}\mathtt{j}}\mathbf{E}^{\mathtt{k}}\wedge\mathbf{E}^{\mathtt{i}}$\tabularnewline
$\frac{1}{\ell}\mathbf{d}\mathbf{E}^{\mathtt{i}}=\frac{1}{\ell}\mathbf{E}^{\mathtt{k}}\wedge\boldsymbol{\omega}_{\;\mathtt{k}}^{\mathtt{i}}$\tabularnewline
\hline 
\end{tabular}
}
\end{center}

\begin{center}
\begin{tabular}{|c|}
\hline 
Conformal\tabularnewline
\hline 
$\boldsymbol{\theta}=\boldsymbol{\omega}_{\;b}^{a}M_{\;a}^{b}+\boldsymbol{\omega}^{a}P_{a}+\boldsymbol{\omega}_{a}K^{a}+\boldsymbol{\omega}D$\tabularnewline
\hline 
$\mathbf{d\boldsymbol{\mathbf{\omega}}}_{\;b}^{a}=\boldsymbol{\mathbf{\omega}}_{\;b}^{c}\wedge\boldsymbol{\mathbf{\omega}}_{\;c}^{a}+2\Delta_{cb}^{ad}\boldsymbol{\omega}_{d}\wedge\boldsymbol{\omega}^{c}$\tabularnewline
$\mathbf{d}\boldsymbol{\omega}^{a}=\boldsymbol{\omega}^{c}\wedge\boldsymbol{\mathbf{\omega}}_{\;c}^{a}+\boldsymbol{\mathbf{\omega}}\wedge\boldsymbol{\omega}^{a}$\tabularnewline
$\mathbf{d}\boldsymbol{\omega}_{a}=-\boldsymbol{\omega}_{b}\wedge\boldsymbol{\mathbf{\omega}}_{\;a}^{b}-\mathbf{\boldsymbol{\omega}}\wedge\boldsymbol{\omega}_{a}$\tabularnewline
$\mathbf{d}\boldsymbol{\omega}=\boldsymbol{\omega}^{a}\wedge\boldsymbol{\omega}_{a}$\tabularnewline
\hline 
\end{tabular}
\end{center}

\egroup
\protect\caption{Maurer-Cartan Equations for $SO\left(p,q\right)$ written in the AdS, psuedo-rotatation and conformal generator bases. The projection $\Delta_{cb}^{ad}\equiv\frac{1}{2}\left(\delta_{c}^{a}\delta_{b}^{d}-\eta^{ad}\eta_{cb}\right)$ preserves the antisymmetry of the spin connection.}\label{MCEqnTable}
\end{table}

Note in the conformal case of Table (\ref{MCEqnTable}) that, in our conventions, the forms $\boldsymbol{\omega}^{a}$ and $\boldsymbol{\omega}_{a}$
are distinct and are related to the translations and special conformal
transformations, respectively. One can move from one version of the equations to another simply
by using the relations between the bases given in Eqn (\ref{eq:LieAlgebraBases}).
These relations will give us the direct correspondence between the forms
in the AdS version of the MC equation and the conformal group version
of the MC equation and can be written down explicitly,
\begin{eqnarray}
\boldsymbol{\omega}_{\;b}^{a} & = & \boldsymbol{\omega}_{\;b}^{a}\nonumber \\
\boldsymbol{\omega}_{\;4}^{a} & = & \frac{1}{\sqrt{2}}\left(\boldsymbol{\omega}^{a}-\eta^{ab}\boldsymbol{\omega}_{b}\right)\nonumber \\
\frac{1}{\ell}\mathbf{E}^{a}=\boldsymbol{\omega}_{\;5}^{a} & = & \frac{1}{\sqrt{2}}\left(\boldsymbol{\omega}^{a}+\eta^{ab}\boldsymbol{\omega}_{b}\right)\nonumber \\
\frac{1}{\ell}\mathbf{E}^{4}=\boldsymbol{\omega}_{\;5}^{4} & = & \boldsymbol{\omega} .\label{eq:Correspondence}
\end{eqnarray}
where the left side lists forms from the AdS basis and the right side
lists the equivalent combination of forms from the conformal basis.
Remember that our nested notation allows us to write the connections
in the following way, $\boldsymbol{\omega}_{\;B}^{A}=\left(\boldsymbol{\omega}_{\;\mathtt{j}}^{\mathtt{i}},\boldsymbol{\omega}_{\;5}^{\mathtt{i}}\right)=\left(\boldsymbol{\omega}_{\;\mathtt{j}}^{\mathtt{i}},\frac{1}{\ell}\mathbf{E}^{\mathtt{i}}\right)=\left(\boldsymbol{\omega}_{\;b}^{a},\boldsymbol{\omega}_{\;4}^{a},\boldsymbol{\omega}_{\;5}^{a},\boldsymbol{\omega}_{\;5}^{4}\right)$,
where we have omitted redundancies from the antisymmetry of the forms.
It is these mathematical relationships that inspired Maldecena's original conjecture. 
Our construction will break the equality of the left and right sides of Eqn~(\ref{eq:Correspondence}), since this only holds in the full group manifold. Once we use Cartan geometry to build an action, the correspondence in Eqn~(\ref{eq:Correspondence}) gives rise to an explicit physical example of the AdS/CFT correspondence.

\section{Building a Physical Theory}

In order to build a physical theory we begin by constructing a Cartan
geometry. The MC equation is the foundation of the geometric information
given by the Lie group structure. The brief introduction below is
in no way comprehensive, for more background see references listed in the introduction.

\subsection{Quotient Manifolds as Homogeneous Model Spaces}

Quotients of the group $SO\left(4,2\right)$ by different subgroups
will allow us to look at theories based on the different local symmetries.
In general, a Lie group, $\mathcal{G}$, quotiented by a subgroup,
$\mathcal{H}$, gives a principal fiber bundle with base manifold,
$\mathcal{M}$, of dimension, $dim\left(\mathcal{M}\right)=dim\left(\mathcal{G}\right)-dim\left(\mathcal{H}\right)$.
The Maurer-Cartan form on $\mathcal{G}$ induces a (flat) Cartan connection
on this quotient (principal fiber bundle) taking values in the Lie
algebra $\mathfrak{g}$ associated with $\mathcal{G}$. The Cartan
construction then uses this homogeneous space as a model space by
looking at the bundle of these spaces over an, in general curved,
manifold of the same dimension as $\mathcal{M}$. 

The Cartan curvature, $\boldsymbol{\Omega}_{\;B}^{A}$, for any $SO\left(p,q\right)$ group can be written
as
\begin{eqnarray*}
\boldsymbol{\Omega}_{\;B}^{A} & = & \mathbf{d}\boldsymbol{\omega}_{\;B}^{A}-\boldsymbol{\omega}_{\;B}^{C}\boldsymbol{\omega}_{\;C}^{A},
\end{eqnarray*}
where the curvature encodes the
deviation from the manifold being flat\footnote{We have abused notation slightly here, since the connections on the curved manifold are not the same as those of the homogeneous manifold, but use the same notation since it is obvious from context.}. The curvatures are required to be horizontal, and thus describe curvature of the manifold only.

Note that all subsequent versions
of the Cartan equation are nothing but different choices of basis
for the generators of the Lie algebra that better suit the local
symmetry of our theory. These curvatures, along with any other tensorial
objects (e.g. the totally anti-symmetric symbol, $\epsilon_{ABCDEF}$)
can then be used to construct an action principle, and hence, a physical
theory.

\subsection{Example: Anti-de Sitter Relativity}

As a pertinent example we take the quotient $SO(4,2)/SO(4,1)$. This
gives us a principal fiber bundle with a 5-dim base manifold, where
we interpret the fibers as the Lorentz symmetries of a 5-dim space. Through
the Cartan construction this homogeneous space is used as the model
space for a curved manifold. The set of structure equations on the
manifold are identical in form to the standard tetrad version of an
Einstein-Cartan geometry with cosmological constant, $\Lambda=-\frac{\left(n-1\right)\left(n-2\right)}{2\ell^{2}}.$
\begin{eqnarray}
\boldsymbol{\Omega}_{\;\mathtt{j}}^{\mathtt{i}} & = & \mathbf{d}\boldsymbol{\omega}_{\;\mathtt{j}}^{\mathtt{i}}-\boldsymbol{\omega}_{\;\mathtt{j}}^{\mathtt{k}}\boldsymbol{\omega}_{\;\mathtt{k}}^{\mathtt{i}}+\frac{1}{\ell^{2}}\eta_{\mathtt{k}\mathtt{j}}\mathbf{E}^{\mathtt{k}}\mathbf{E}^{\mathtt{i}}\label{eq:AdS Curvature}\\
\frac{1}{\ell}\boldsymbol{\mathtt{T}}^{\mathtt{i}} & = & \frac{1}{\ell}\mathbf{d}\mathbf{E}^{\mathtt{i}}-\frac{1}{\ell}\mathbf{E}^{\mathtt{k}}\boldsymbol{\omega}_{\;\mathtt{k}}^{\mathtt{i}}\label{eq:AdS Torsion}
\end{eqnarray}
We have included the torsion, $\boldsymbol{\mathtt{T}}^{\mathtt{i}}$,
for full generality, but, just as in Einstein-Cartan theory, our action
will set $\mathtt{T}_{\;\mathtt{j}\mathtt{k}}^{\mathtt{i}}=0$. Note
that the Riemann curvature 2-form, $\mathbf{R}_{\;\mathtt{j}}^{\mathtt{i}}\equiv\mathbf{d}\boldsymbol{\omega}_{\;\mathtt{j}}^{\mathtt{i}}-\boldsymbol{\omega}_{\;\mathtt{j}}^{\mathtt{k}}\boldsymbol{\omega}_{\;\mathtt{k}}^{\mathtt{i}}$ is only a part of the full Cartan curvature of the Lorentz connection, $\boldsymbol{\Omega}_{\;\mathtt{j}}^{\mathtt{i}}$.
This geometry has an extra term, $\frac{1}{\ell^{2}}\eta_{\mathtt{k}\mathtt{j}}\mathbf{E}^{\mathtt{k}}\mathbf{E}^{\mathtt{i}}$, that accounts for the full scope of the AdS symmetry.
It is especially important to note that a zero curvature solution $\left(\boldsymbol{\Omega}_{\;\mathtt{j}}^{\mathtt{i}}=0\right)$ will be equivalent to the 5-dimensional AdS solution of general relativity, and acts as the equivalent to the Minkowski tangent spaces of general relativity.
More general solutions will not have the full AdS symmetry, but are
locally symmetric, inline with the flat version of Eqn~(\ref{eq:AdS Curvature}); just as general solutions of general relativity with vanishing $\Lambda$
do not have Poincar\'{e} symmetry, but are locally Lorentz symmetric.

With this curvature we can write down an action, using differential forms, that is identical
to the Einstein-Cartan action, $S=\int\eta^{\mathtt{i}\mathtt{n}}\boldsymbol{\Omega}_{\;\mathtt{n}}^{\mathtt{j}}{}^{\wedge}\mathbf{E}^{\mathtt{k}}{}^{\wedge}\mathbf{E}^{\mathtt{l}}{}^{\wedge}\mathbf{E}^{\mathtt{m}}{}\epsilon_{\mathtt{i}\mathtt{j}\mathtt{k}\mathtt{l}\mathtt{m}}$, except that we use the Cartan 
curvature in $5$-dim, instead of the Riemann curvature in $4$-dim.
Varying the action with respect to all the connections (a Palatini
variation) gives the expected field equations. 
\begin{eqnarray*}
\Omega_{\mathtt{i}\mathtt{j}}-\frac{1}{2}\Omega\eta_{\mathtt{i}\mathtt{j}} & = & 0\Rightarrow\Omega_{\mathtt{i}\mathtt{j}}=0\\
\mathtt{T}_{\;\mathtt{j}\mathtt{k}}^{\mathtt{i}} & = & 0
\end{eqnarray*}
where $\Omega_{\mathtt{i}\mathtt{j}}=\Omega_{\;\mathtt{i}\mathtt{k}\mathtt{j}}^{\mathtt{k}}$
and $\Omega=\eta^{\mathtt{i}\mathtt{j}}\Omega_{\;\mathtt{i}\mathtt{k}\mathtt{j}}^{\mathtt{k}}$.
The metric, defined as the restriction of the Killing form of $SO\left(4,2\right)$
to the base manifold, is used to define the inner product of the basis
forms, $\left\langle \mathbf{E}^{\mathtt{i}},\mathbf{E}^{\mathtt{j}}\right\rangle =\ell^2 \eta^{\mathtt{i}\mathtt{j}}$. 
Unlike in a Riemannian geometry, the metric comes directly from the symmetry structures. 
Written in terms of the Riemann curvature and the cosmological constant,
$\Omega_{\mathtt{j}\mathtt{l}}=R_{\mathtt{j}\mathtt{l}}-\frac{2\Lambda}{\left(n-2\right)}\eta_{\mathtt{l}\mathtt{j}}$,
we have the vacuum Einstein equation with cosmological constant, 
\begin{eqnarray*}
R_{\mathtt{i}\mathtt{j}} & = & -\frac{2}{\left(n-2\right)}\Lambda\eta_{\mathtt{i}\mathtt{j}}.
\end{eqnarray*}
These equations are identical to the Einstein equation with negative
cosmological constant. We have constructed a theory whose flat solution,
in the sense of the Cartan curvature, is $AdS_{5}$ and whose vacuum solution always possesses a cosmological constant. This result holds in any dimension
$n>2$. One only needs to start with $\mathcal{G}=SO(5,1)$ to have the de~Sitter case.

\subsection{Lie Algebra-Valued Curvature}

One final note about these theories is that there is a choice in writing
the action. The action can be written with the full $\mathcal{G}$-symmetry
and then projected to a space with the more restricted symmetry.
This is the choice used for the familiar MacDowell-Mansouri action
\cite{MacDowell:1977p1566,Stelle:1979aj,Wise:2006sm}. In \cite{Westman:2012zk} Westman and Z\l{}o\'{s}nik use this type of approach  to investigate the AdS/CFT correspondence. Above, however,
we have chosen to build the action using tensors of the more restricted
local $\mathcal{H}$-symmetry of the quotient manifold.

Regardless of this choice, the curvature $2$-forms always take values
in the Lie algebra $\mathfrak{g}$, giving immediate correspondences
between them: 
\begin{eqnarray} 
\boldsymbol{\Omega}_{\;b}^{a} & \Leftrightarrow & \boldsymbol{\Omega}_{\;b}^{a}\nonumber \\
\mathbf{T}^{a} & \Leftrightarrow & \frac{1}{\sqrt{2}}\left(\boldsymbol{\Omega}_{\;5}^{a}+\boldsymbol{\Omega}_{\;4}^{a}\right)=\frac{1}{\sqrt{2}}\left(\frac{1}{\ell}\boldsymbol{\mathtt{T}}^{a}+\boldsymbol{\Omega}_{\;4}^{a}\right)\nonumber \\
\mathbf{S}_{a} & \Leftrightarrow & \frac{1}{\sqrt{2}}\eta_{ab}\left(\boldsymbol{\Omega}_{\;5}^{b}-\boldsymbol{\Omega}_{\;4}^{b}\right)\nonumber \\
\boldsymbol{\Omega} & \Leftrightarrow & \boldsymbol{\Omega}_{\;5}^{4}=\frac{1}{\ell}\boldsymbol{\mathtt{T}}^{4} \label{eq:CurvatureCorres}
\end{eqnarray}
Although these curvatures are horizontal with respect to different
fiberings of $\mathcal{G}$, we may regard them all as restrictions
of Cartan curvatures of the full group, $\mathcal{M}=\mathcal{G}/\mathbb{I}=\mathcal{G}$.
Thus, each is a special case of
\[
\mathbf{d}\boldsymbol{\omega}^{A}=-\frac{1}{2}c_{\;\;BC}^{A}\boldsymbol{\omega}^{B}{}^{\wedge}\boldsymbol{\omega}^{C}+\boldsymbol{\Omega}^{A}
\]
where $\boldsymbol{\Omega}^{A}=\frac{1}{2}\boldsymbol{\Omega}_{\;BC}^{A}\boldsymbol{\omega}^{B}\wedge\boldsymbol{\omega}^{C}$.

\section{Translation in AdS is the Weyl Connection in the CFT}

The purpose of this section is to demonstrate how the methods used
in the standard AdS/CFT correspondence connect with the links seen through
the lens of Cartan geometry. In $\left(n+1\right)$-dim one may write
down the AdS metric in conformal coordinates as 
\begin{eqnarray}
ds^{2} & = & \frac{\ell^{2}}{r^{2}}dr^{2}+\frac{\ell^{2}}{r^{2}}\left(dx^{\mu}dx^{\nu}\eta_{\mu\nu}\right),\label{eq:AdS Metric}
\end{eqnarray}
where $\eta_{\mu\nu}$ is the $n-$dim Minkowski metric. The process
by which one corresponds an $\left(n+1\right)$-dim gravitational
theory to an $n$-dimensional conformal field theory in the context of string theory relies on the
fact that in the limit $r\rightarrow\infty$ the boundary of the ``bulk''
metric appears as an $n$-dim conformally invariant theory based on
Minkowski space. As stated in \cite{Hubeny:2014bla} ``the gauge
theory can be naturally thought of as `living on the boundary' of
AdS: it is formulated on a spacetime which is in the same conformal
class (the relevant structure for a conformal field theory) as the
boundary metric induced from the bulk''.

Remarkably, when we look at the correspondence given in Eqn (\ref{eq:Correspondence}),
we see that one of the $1$-forms spanning the co-tangent space of
the $5$-dim manifold (here chosen as $\mathbf{E}^{4}$) corresponds to the Weyl connection
in the conformal theory, the connection related to the generator of
dilatational symmetries. In other words, the $r$ coordinate direction
becomes the conformal scaling freedom of the metric, obvious from
Eqn (\ref{eq:AdS Metric}) and from the correspondence given in Eqn
(\ref{eq:Correspondence}). Explicitly we can set
\begin{eqnarray*}
\mathbf{E}^{a} & = & \frac{\ell}{r}\delta_{\mu}^{a}\mathbf{d}x^{\mu}\\
\mathbf{E}^{4} & = & \frac{\ell}{r}\mathbf{d}r
\end{eqnarray*}
to get an orthonormal basis. Then the correspondence $\mathbf{E}^{4}=\boldsymbol{\omega}$
shows that
\begin{eqnarray*}
\boldsymbol{\omega} & = & \frac{1}{r}\mathbf{d}r=\mathbf{d}\left(\ln r\right)
\end{eqnarray*}
and a gauge change $e^{\phi}$ that changes the Weyl vector by $\mathbf{d}\phi=\mathbf{d}\left(\ln r\right)$
means that $e^{\phi}=r$.

\section{Two Conformal Cases }

There are four distinct Cartan equations when written in the conformal generator basis. They relate the connection $1$-forms of Lorentz transformations, translations, special conformal transformations
and dilatations and their derivatives. 
\begin{eqnarray*}
\mathbf{d\boldsymbol{\mathbf{\omega}}}_{\;b}^{a} & = & \boldsymbol{\mathbf{\omega}}_{\;b}^{c}\boldsymbol{\mathbf{\omega}}_{\;c}^{a}+2\Delta_{cb}^{ad}\boldsymbol{\omega}_{d}\boldsymbol{\omega}^{c}+\boldsymbol{\Omega}_{\;b}^{a}\\
\mathbf{d}\boldsymbol{\omega}^{a} & = & \boldsymbol{\omega}^{c}\boldsymbol{\mathbf{\omega}}_{\;c}^{a}+\boldsymbol{\mathbf{\omega}}\boldsymbol{\omega}^{a}+\mathbf{T}^{a}\\
\mathbf{d}\boldsymbol{\omega}_{a} & = & -\boldsymbol{\omega}_{b}\boldsymbol{\mathbf{\omega}}_{\;a}^{b}-\mathbf{\boldsymbol{\omega}}\boldsymbol{\omega}_{a}+\mathbf{S}_{a}\\
\mathbf{d}\boldsymbol{\omega} & = & \boldsymbol{\omega}^{a}\boldsymbol{\omega}_{a}+\boldsymbol{\Omega}
\end{eqnarray*}

Of the scale invariant quotients, there are two options to choose
from that keep Lorentz symmetry: $\mathcal{H}$ can either be the Weyl group (Lorentz group with
dilatations) or the inhomogeneous Weyl group, (Weyl group plus translations)\footnote{There are two further choices that keep Lorentz symmetry on the fibers, but these break the scaling symmetry}.

\subsection{The Auxiliary (Parabolic) Case}

The quotient of $SO\left(4,2\right)$ by the inhomogeneous Weyl group
puts Lorentz transformations, dilatations and the special conformal
transformations on the fibers. This leaves a 4-dim base manifold that
is locally symmetric under the inhomogeneous Weyl group. Most conformally symmetric
gravity theories are based on this set of symmetries being on the
fibers of the bundle, including Weyl gravity, Tractor calculus and
the boundary of AdS space in general relativity. The projection of the Killing metric to the base manifold in this quotient is degenerate, therefore it cannot be used as an inner product. 
We mention this case, only to contrast it with the biconformal case discussed in the next subsection that has a non-degenerate projection and hence an inner product. 

In addition, the $\boldsymbol{\omega}_{a}$ have been shown to be auxiliary, in the sense
that no matter which Lagrangian is chosen the $\boldsymbol{\omega}_{a}$ can always be written in terms of the Riemann curvature and its traces \cite{Wheeler:1991ff}, see also \cite{Attard:2015jfr}.

\subsection{The Biconformal Case}

The quotient $SO\left(4,2\right)/\left(SO\left(3,1\right)\times SO\left(1,1\right)\right)$,
where we refer to the quotient manifold (and its curved generalization)
as biconformal space (BCS), has been studied in detail by the author and
collaborators. Most often the ``extra'' $4$-dim\footnote{In fact, calculations in biconformal space are often straightforward
to do for all groups of $Conf\left(p,q\right)$ and $n=p+q.$}, spanned by the $\boldsymbol{\omega}_{a}$ fields, have been interpreted
as a co-tangent space to the other $4$-dim, hence those dimensions
are thought of as momenta in a particle theory \cite{Wheeler:1994an,Wheeler:1997pc,Wheeler:1997tg,Wehner:1999p1653,Hazboun:2013lra,Anderson:2004yh}.
This stems from an interpretation of the the dilatation structure
equation as a symplectic form, i.e. a closed non-degenerate $2$-form
that spans the base manifold, and sets up a canonical structure. In a Cartan geometry constructed from this quotient the dynamics are not restricted as in the auxiliary case. Therefore, the
most important physical difference between biconformal space and the
auxiliary case is that in this case one can construct a Lagrangian,
linear in the curvatures, which is conformally invariant \cite{Wehner:1999p1653}. 
The field equations manifest the Einstein field equation in most simplifications. While the vacuum Einstein equation still appears, the main result of this manuscript is based instead on interpreting the $\boldsymbol{\omega}_{a}$ as gauge fields on spacetime.

There are two options for the inner product of the two submanifolds of biconformal space. One can endow the submanifolds each with the inner-product of their choosing (usually Lorentzian), as one does in a Poincar\'e gauge theory of gravity, (and in fact most constructions of GR). However this is not necessary in biconformal space where the group structure hands one a metric. Unlike the other conformally symmetric quotient, the Killing metric of the generators projected to the quotient manifold in biconformal space is non-degenerate.
An explicit basis for the connections of homogeneous biconformal space is provided by Wheeler,~\cite{Wheeler:1994an}, 
\begin{eqnarray*}
\boldsymbol{\omega}_{\;b}^{a} & = & \left(\delta_{c}^{a}\delta_{b}^{d}-\eta^{ad}\eta_{cb}\right)s_{d}\mathbf{d}w^{c}\\
\boldsymbol{\omega}^{a} & = & \mathbf{d}w^{a}\\
\boldsymbol{\omega}_{a} & = & \mathbf{d}s_{a}+b_{ab}\mathbf{d}w^{b}=\mathbf{d}s_{a}-\left(s_{a}s_{b}-\frac{1}{2}s^{2}\eta_{ab}\right)\mathbf{d}w^{b}\\
\boldsymbol{\omega} & = & -s_{a}\mathbf{d}w^{a} \label{eq:FlatBCS}
\end{eqnarray*}
where $w^{a}$ are coordinates spanning the spacetime.
The other $4$-dim have coordinates $s_{a}$, and possess an inner product, derived from
the Killing metric.

\begin{eqnarray*}
\left\langle \mathbf{d}s_{a},\mathbf{d}s_{b}\right\rangle &=&\left\langle \boldsymbol{\omega}_{a}+\left(s_{a}s_{c}-\frac{1}{2}s^{2}\eta_{ac}\right)\mathbf{d}w^{c},\boldsymbol{\omega}_{b}+\left(s_{b}s_{d}-\frac{1}{2}s^{2}\eta_{bd}\right)\mathbf{d}w^{d}\right\rangle \\&=&\left(s_{b}s_{d}-\frac{1}{2}s^{2}\eta_{bd}\right)\delta_{a}^{d}+\left(s_{a}s_{c}-\frac{1}{2}s^{2}\eta_{ac}\right)\delta_{b}^{c}\\&=&\left(2s_{a}s_{b}-s^{2}\eta_{ab}\right)\equiv2b_{ab}
\end{eqnarray*}

Just as in Riemannian geometry, where a curved geometry may inherit an inner product from the tangent space, a Cartan geometry may inherit an inner product from the homogeneous space. It has been shown in the $SO(5,1)$ case \cite{Spencer:2008p167} that the form of this metric changes the signature of the original metric from a positive-definite Euclidean inner product to one with a Lorentzian signature, $\eta_{ab}=diag\left(-1,1,1,1\right)$.  As shown in \cite{Spencer:2008p167}, if one wants complementary Lagrangian submanifolds they necessarily have $(p, q)$ and $(q, p)$ signatures, respectively. However, if the submanifolds are oblique it is possible to place the Minkowski metric on one and Euclidean metric on the other without imposing anything by hand. The fact that the separate submanifolds of biconformal space may possess either a Euclidean and/or Lorentzian signature metric is crucial to the conclusion of this manuscript.

\section{Discussion}

The goal of this note is to demonstrate how a Cartan geometry gives the correspondence between physical theories based on the same overarching symmetry group,~$\mathcal{G}$. The correspondence of the connections of the Cartan geometry can be viewed from a different geometric perspective by noting that the $SO(4, 1)$ subgroup may be achieved by holding one of the time dimensions of $SO(4, 2)$ fixed. The biconformal symmetry may be seen geometrically as holding one spatial dimension of the $SO(4, 1)$ symmetry apart from the rest, but allowing boosts between it and the fixed time direction that gives the $SO(4,1)$. Therefore, the $SO(3, 1)$ subgroups may be exactly identified between the two quotients. 

It remains to describe where the degrees of freedom of a Yang-Mills theory emerge. There are a number of candidates already extant in the literature of a (conformal) Yang-Mills theory that can be constructed on the biconformal space. It is out of the scope here to cover them in detail, however below we summarize two possible routes to the YM theory.

Imposing the Lorentzian signature metrics on both submanifolds of BCS, a straightforward route to an $SU(2)$ YM theory is via a graviweak theory \cite{Nesti:2007ka}. In the construction by Nesti and Percacci, the anti-selfdual sub-algebra of the (complexified) $SO(1,3)$ symmetry is identified with the $SU(2)_L$ isospin group appearing in the standard model. This decomposition allows for the building of a fermionic action, and in the case of biconformal space the theory is conformally symmetric by virtue of the Weyl connection. Another example of such a theory, using an extended Plebanski action, is discussed in~\cite{Das:2013xha}. Such a theory's degrees of freedom would use the ``extra'' connections of BCS or be a part of the 4-dimensional theory in the parabolic case. 

The existence of a Euclidean inner product in BCS allows for an even cleaner decomposition, since $SO(4)$ is isomorphic to $SU(2)\times SU(2)$. Such a decomposition is well known in the Euclidean gravity community from research on gravitational instantons \cite{Eguchi:1979yx}. Using a similar formalism to \cite{Oh:2011nv}, one can decompose the Riemann tensor into self-dual and anti self-dual parts.  
\begin{equation} \label{eq:SU2_decomp}
R_{abcd}=F_{ab}^{(+)A}T_{cd}^{(+)A}+F_{ab}^{(-)A}T_{cd}^{(-)A}
\end{equation}
The $F_{ab}^{(\pm)A}$ are $SU(2)$ curvatures, the self-dual ($+$) and anti self-dual ($-$) components of the $SO(4)$ curvature, the $T_{cd}^{(\pm)A}$ are $4\times4$ matrix representations of $SU(2)$ generators and the capital $A$ indices run $1,2,3$, over the two sets of $SU(2)$ generators. The lower case latin indices are in the usual orthonormal frame. 
Decomposing the Riemann tensor into (anti)self-dual pieces, as in Eqn~(\ref{eq:SU2_decomp}), and taking the divergence gives
\begin{equation} \label{eq:divergenec_SU2_decomp}
D^{a}R_{abcd}=D^{a}F_{ab}^{(+)A}T_{cd}^{(+)A}+D^{a}F_{ab}^{(-)A}T_{cd}^{(-)A}.
\end{equation}
If the metric is self-dual or anti self-dual (i.e one of the $F_{ab}^{(\pm)A}=0$) then the theory is equivalent to a topological $SU(2)$ YM theory\footnote{This problem has also be cast in the language of gauge theory using the Ashtekar connection by Samuel \cite{Samuel:1988jx}.}, the instanton. This follows from the triple antisymmetry of the 2nd Bianchi identity. See \cite{Oh:2011nv} for details.  A generic metric with a curvature tensor possessing both (anti)self-dual pieces is not trivial, as is the instanton, but still decomposable. In this more general case the divergence of the Riemann tensor can be replaced by the curl of the Ricci tensor, which is zero for a large class of solutions to the Einstein field equation, such as Einstein manifolds. Then Eqn~(\ref{eq:divergenec_SU2_decomp}) is equivalent to two vacuum YM field equations. If the curl of the Ricci tensor does not vanish then it can be decomposed into (anti)self-dual components and viewed as a source for the YM fields. This association, known since the early 1980s in the context of Euclidean gravity, sets up biconformal space as the perfect stage for a AdS/CFT correspondence, where one submanifold can be interpreted as the background spacetime, and one submanifold, along with its connections, can act as the CFT. 

Using this ``CFT from Euclidean Gravity'' approach it is straightforward to track the connections through Eqns~(\ref{eq:Correspondence}),  (\ref{eq:CurvatureCorres}) \& (\ref{eq:FlatBCS}) from the anti-de~Sitter theory through to the effective YM theory on biconformal space. Since the theory is manifestly conformally symmetric one obtains a conformally symmetric Yang-Mills field.

A different approach to building a particle theory from gravitational degrees of freedom in BCS is studied by Lovelady and Wheeler \cite{Lovelady:2015xhh}, where they investigate the $SO(5,1)$ (de~Sitter) case. Their work shows the broad applicability of Cartan geometry to these types of problems. In fact, it should be possible to show, using the relationships between the BMS symmetry group and the Galilean conformal algebra explored by Bagchi, et al. \cite{Bagchi:2010,Bagchi:2013toa}, how to use a quotient of the Poincar\'e group to track a similar correspondence using Cartan geometry. 

We have shown that building two theories, one with AdS symmetry on
a 5-dimensional manifold and one with conformal symmetry on a 4-dimensional
manifold is straight forward in the context of Cartan geometry. The correspondence we speak of here is not the AdS/CFT correspondence
of string theory. The main difference is that we do not show a connection
between a \emph{classical }gravitational theory and a \emph{quantum
} Yang-Mills theory. Instead we show that the classical
theories based on AdS or conformal symmetry have a natural correspondence
in the context of their Lie algebras alone. For this correspondence, we have shown that neither SUSY, nor holography is necessary.

\section*{Acknowledgements}
This work was supported in part by the NSF Physics Frontier Center award number 1430284. The author wishes to thank B\'eatrice Bonga and James T. Wheeler for discussions about this subject and looking through earlier versions of the draft. I would also like to thank Derek Wise for a private communication concerning Cartan curvatures and Charles Torre for discussions about gravitational instantons.

\bibliographystyle{elsarticle-num}%iopart-num} 

\bibliography{AdSCFTBib}

\begin{thebibliography}{10}
\expandafter\ifx\csname url\endcsname\relax
  \def\url#1{\texttt{#1}}\fi
\expandafter\ifx\csname urlprefix\endcsname\relax\def\urlprefix{URL }\fi
\expandafter\ifx\csname href\endcsname\relax
  \def\href#1#2{#2} \def\path#1{#1}\fi

\bibitem{Maldacena:1997re}
J.~M. Maldacena, {The Large N limit of superconformal field theories and
  supergravity}, Int.J.Theor.Phys. 38 (1999) 1113--1133.
\newblock \href {http://arxiv.org/abs/hep-th/9711200}
  {\path{arXiv:hep-th/9711200}}, \href
  {http://dx.doi.org/10.1023/A:1026654312961}
  {\path{doi:10.1023/A:1026654312961}}.

\bibitem{Hartnoll:2008kx}
S.~A. Hartnoll, C.~P. Herzog, G.~T. Horowitz, {Holographic Superconductors},
  JHEP 12 (2008) 015.
\newblock \href {http://arxiv.org/abs/0810.1563} {\path{arXiv:0810.1563}},
  \href {http://dx.doi.org/10.1088/1126-6708/2008/12/015}
  {\path{doi:10.1088/1126-6708/2008/12/015}}.

\bibitem{Erlich:2005qh}
J.~Erlich, E.~Katz, D.~T. Son, M.~A. Stephanov, {QCD and a holographic model of
  hadrons}, Phys. Rev. Lett. 95 (2005) 261602.
\newblock \href {http://arxiv.org/abs/hep-ph/0501128}
  {\path{arXiv:hep-ph/0501128}}, \href
  {http://dx.doi.org/10.1103/PhysRevLett.95.261602}
  {\path{doi:10.1103/PhysRevLett.95.261602}}.

\bibitem{Attard:2015jfr}
J.~Attard, J.~Fran{\c c}ois, S.~Lazzarini, {Weyl gravity and Cartan geometry},
  Phys. Rev. D93~(8) (2016) 085032.
\newblock \href {http://arxiv.org/abs/1512.06907} {\path{arXiv:1512.06907}},
  \href {http://dx.doi.org/10.1103/PhysRevD.93.085032}
  {\path{doi:10.1103/PhysRevD.93.085032}}.

\bibitem{Nesti:2007ka}
F.~Nesti, R.~Percacci, {Graviweak Unification}, J. Phys. A41 (2008) 075405.
\newblock \href {http://arxiv.org/abs/0706.3307} {\path{arXiv:0706.3307}},
  \href {http://dx.doi.org/10.1088/1751-8113/41/7/075405}
  {\path{doi:10.1088/1751-8113/41/7/075405}}.

\bibitem{Eguchi:1979yx}
T.~Eguchi, A.~J. Hanson, {Gravitational Instantons}, Gen. Rel. Grav. 11 (1979)
  315--320.
\newblock \href {http://dx.doi.org/10.1007/BF00759271}
  {\path{doi:10.1007/BF00759271}}.

\bibitem{Wheeler:1997pc}
J.~T. Wheeler, {New conformal gauging and the electromagnetic theory of Weyl},
  J.Math.Phys. 39 (1998) 299--328.
\newblock \href {http://arxiv.org/abs/hep-th/9706214}
  {\path{arXiv:hep-th/9706214}}, \href {http://dx.doi.org/10.1063/1.532315}
  {\path{doi:10.1063/1.532315}}.

\bibitem{Sharpe:1997a}
R.~W. Sharpe, Differential geometry, Vol. 166 of Graduate Texts in Mathematics,
  Springer-Verlag, New York, 1997.

\bibitem{Cartan1910a}
E.~Cartan, \href{http://www.numdam.org/item?id=ASENS_1910_3_27__109_0}{Les
  syst\`emes de {P}faff, \`a cinq variables et les \'equations aux d\'eriv\'ees
  partielles du second ordre}, Ann. Sci. \'Ecole Norm. Sup. (3) 27 (1910)
  109--192.
\newline\urlprefix\url{http://www.numdam.org/item?id=ASENS_1910_3_27__109_0}

\bibitem{Utiyama:1956p1235}
R.~Utiyama, \href{http://prola.aps.org/abstract/PR/v101/i5/p1597\_1}{Invariant
  theoretical interpretation of interaction}, Physical Review 101 (1956)
  1597--1607.
\newline\urlprefix\url{http://prola.aps.org/abstract/PR/v101/i5/p1597\_1}

\bibitem{Kibble:1961p1468}
T.~Kibble, \href{http://jmp.aip.org/resource/1/jmapaq/v2/i2/p212\_s1}{{Lorentz
  invariance and the gravitational field}}, J.Math.Phys. 2 (1961) 212--221.
\newline\urlprefix\url{http://jmp.aip.org/resource/1/jmapaq/v2/i2/p212\_s1}

\bibitem{MacDowell:1977p1566}
S.~W. MacDowell, F.~Mansouri,
  \href{http://link.aps.org/doi/10.1103/PhysRevLett.38.739}{Unified geometric
  theory of gravity and supergravity}, Phys. Rev. Lett. 38 (1977) 739--742.
\newblock \href {http://dx.doi.org/10.1103/PhysRevLett.38.739}
  {\path{doi:10.1103/PhysRevLett.38.739}}.
\newline\urlprefix\url{http://link.aps.org/doi/10.1103/PhysRevLett.38.739}

\bibitem{Neeman:1978p1517}
Y.~Ne'eman, T.~Regge,
  \href{http://www.sciencedirect.com/science/article/pii/0370269378900588}{Gravity
  and supergravity as gauge theories on a group manifold}, Physics Letters B 74
  (1978) 54 -- 56.
\newblock \href {http://dx.doi.org/10.1016/0370-2693(78)90058-8}
  {\path{doi:10.1016/0370-2693(78)90058-8}}.
\newline\urlprefix\url{http://www.sciencedirect.com/science/article/pii/0370269378900588}

\bibitem{Neeman:1978p1521}
Y.~Ne'eman, T.~Regge, \href{http://dx.doi.org/10.1007/BF02724472}{Gauge theory
  of gravity and supergravity on a group manifold}, La Rivista del Nuovo
  Cimento (1978-1999) 1 (1978) 1--43, 10.1007/BF02724472.
\newline\urlprefix\url{http://dx.doi.org/10.1007/BF02724472}

\bibitem{Ivanov:1982p1172}
E.~A. Ivanov, J.~Niederle,
  \href{http://link.aps.org/doi/10.1103/PhysRevD.25.976}{Gauge formulation of
  gravitation theories. i. the poincar\'e, de sitter, and conformal cases},
  Phys. Rev. D 25 (1982) 976--987.
\newblock \href {http://dx.doi.org/10.1103/PhysRevD.25.976}
  {\path{doi:10.1103/PhysRevD.25.976}}.
\newline\urlprefix\url{http://link.aps.org/doi/10.1103/PhysRevD.25.976}

\bibitem{Ivanov:1982p1201}
E.~A. Ivanov, J.~Niederle,
  \href{http://link.aps.org/doi/10.1103/PhysRevD.25.988}{Gauge formulation of
  gravitation theories. ii. the special conformal case}, Phys. Rev. D 25 (1982)
  988--994.
\newblock \href {http://dx.doi.org/10.1103/PhysRevD.25.988}
  {\path{doi:10.1103/PhysRevD.25.988}}.
\newline\urlprefix\url{http://link.aps.org/doi/10.1103/PhysRevD.25.988}

\bibitem{Hazboun:2013lra}
J.~S. Hazboun, J.~T. Wheeler, {Time and dark matter from the conformal
  symmetries of Euclidean space}, Class. Quant. Grav. 31~(21) (2014) 215001.
\newblock \href {http://arxiv.org/abs/1305.6972} {\path{arXiv:1305.6972}},
  \href {http://dx.doi.org/10.1088/0264-9381/31/21/215001}
  {\path{doi:10.1088/0264-9381/31/21/215001}}.

\bibitem{Westman:2012zk}
H.~F. Westman, T.~G. Zlosnik, {Cartan gravity, matter fields, and the gauge
  principle}, Annals Phys. 334 (2013) 157--197.
\newblock \href {http://arxiv.org/abs/1209.5358} {\path{arXiv:1209.5358}},
  \href {http://dx.doi.org/10.1016/j.aop.2013.03.012}
  {\path{doi:10.1016/j.aop.2013.03.012}}.

\bibitem{Wise:2006sm}
D.~K. Wise, {MacDowell-Mansouri gravity and Cartan geometry}, Class.Quant.Grav.
  27 (2010) 155010.
\newblock \href {http://arxiv.org/abs/gr-qc/0611154}
  {\path{arXiv:gr-qc/0611154}}, \href
  {http://dx.doi.org/10.1088/0264-9381/27/15/155010}
  {\path{doi:10.1088/0264-9381/27/15/155010}}.

\bibitem{Lovelady:2015xhh}
B.~C. Lovelady, J.~T. Wheeler, {Dynamical spacetime symmetry}, Phys. Rev.
  D93~(8) (2016) 085002.
\newblock \href {http://arxiv.org/abs/1512.01729} {\path{arXiv:1512.01729}},
  \href {http://dx.doi.org/10.1103/PhysRevD.93.085002}
  {\path{doi:10.1103/PhysRevD.93.085002}}.

\bibitem{Anderson:2003db}
L.~B. Anderson, J.~T. Wheeler, {Biconformal supergravity and the AdS / CFT
  conjecture}, Nucl.Phys. B686 (2004) 285--309.
\newblock \href {http://arxiv.org/abs/hep-th/0309111}
  {\path{arXiv:hep-th/0309111}}, \href
  {http://dx.doi.org/10.1016/j.nuclphysb.2004.02.034}
  {\path{doi:10.1016/j.nuclphysb.2004.02.034}}.

\bibitem{Stelle:1979aj}
K.~Stelle, P.~C. West, {Spontaneously Broken De Sitter Symmetry and the
  Gravitational Holonomy Group}, Phys.Rev. D21 (1980) 1466.
\newblock \href {http://dx.doi.org/10.1103/PhysRevD.21.1466}
  {\path{doi:10.1103/PhysRevD.21.1466}}.

\bibitem{Hubeny:2014bla}
V.~E. Hubeny, {The AdS/CFT Correspondence}, Class. Quant. Grav. 32~(12) (2015)
  124010.
\newblock \href {http://arxiv.org/abs/1501.00007} {\path{arXiv:1501.00007}},
  \href {http://dx.doi.org/10.1088/0264-9381/32/12/124010}
  {\path{doi:10.1088/0264-9381/32/12/124010}}.

\bibitem{Wheeler:1991ff}
J.~T. Wheeler, {The Auxiliary field in conformal gauge theory}, Phys.Rev. D44
  (1991) 1769--1773.
\newblock \href {http://dx.doi.org/10.1103/PhysRevD.44.1769}
  {\path{doi:10.1103/PhysRevD.44.1769}}.

\bibitem{Wheeler:1994an}
J.~T. Wheeler, \href{http://alice.cern.ch/format/showfull?sysnb=0191411}{{Scale
  invariant phase space and the conformal group}}, in: {On recent developments
  in theoretical and experimental general relativity, gravitation, and
  relativistic field theories. Proceedings, 7th Marcel Grossmann Meeting,
  Stanford, USA, July 24-30, 1994. Pt. A + B}, 1994, pp. 457--459.
\newblock \href {http://arxiv.org/abs/gr-qc/9411030}
  {\path{arXiv:gr-qc/9411030}}.
\newline\urlprefix\url{http://alice.cern.ch/format/showfull?sysnb=0191411}

\bibitem{Wheeler:1997tg}
J.~T. Wheeler, {Why quantum mechanics is complex}, Bull. Astron. Soc. India 25
  (1997) 591.
\newblock \href {http://arxiv.org/abs/hep-th/9708088}
  {\path{arXiv:hep-th/9708088}}.

\bibitem{Wehner:1999p1653}
A.~Wehner, J.~T. Wheeler,
  \href{http://www.sciencedirect.com/science/article/pii/S0550321399003673}{Conformal
  actions in any dimension}, Nuclear Physics B 557 (1999) 380 -- 406.
\newblock \href {http://dx.doi.org/10.1016/S0550-3213(99)00367-3}
  {\path{doi:10.1016/S0550-3213(99)00367-3}}.
\newline\urlprefix\url{http://www.sciencedirect.com/science/article/pii/S0550321399003673}

\bibitem{Anderson:2004yh}
L.~B. Anderson, J.~T. Wheeler, {Quantum mechanics as a measurement theory on
  biconformal space}, Int. J. Geom. Meth. Mod. Phys. 3 (2006) 315--340.
\newblock \href {http://arxiv.org/abs/hep-th/0406159}
  {\path{arXiv:hep-th/0406159}}, \href
  {http://dx.doi.org/10.1142/S0219887806001168}
  {\path{doi:10.1142/S0219887806001168}}.

\bibitem{Spencer:2008p167}
J.~A. Spencer, J.~T. Wheeler, {The existence of time},
  Int.J.Geom.Meth.Mod.Phys. 8 (2011) 273--301.
\newblock \href {http://arxiv.org/abs/0811.0112} {\path{arXiv:0811.0112}},
  \href {http://dx.doi.org/10.1142/S0219887811005130}
  {\path{doi:10.1142/S0219887811005130}}.

\bibitem{Das:2013xha}
C.~R. Das, L.~V. Laperashvili, A.~Tureanu, {Graviweak Unification, Invisible
  Universe and Dark Energy}, Int. J. Mod. Phys. A28 (2013) 1350085.
\newblock \href {http://arxiv.org/abs/1304.3069} {\path{arXiv:1304.3069}},
  \href {http://dx.doi.org/10.1142/S0217751X13500851}
  {\path{doi:10.1142/S0217751X13500851}}.

\bibitem{Oh:2011nv}
J.~J. Oh, C.~Park, H.~S. Yang, {Yang-Mills Instantons from Gravitational
  Instantons}, JHEP 04 (2011) 087.
\newblock \href {http://arxiv.org/abs/1101.1357} {\path{arXiv:1101.1357}},
  \href {http://dx.doi.org/10.1007/JHEP04(2011)087}
  {\path{doi:10.1007/JHEP04(2011)087}}.

\bibitem{Samuel:1988jx}
J.~Samuel, {Gravitational Instantons From the Ashtekar Variables}, Class.
  Quant. Grav. 5 (1988) L123--L125.
\newblock \href {http://dx.doi.org/10.1088/0264-9381/5/8/002}
  {\path{doi:10.1088/0264-9381/5/8/002}}.

\bibitem{Bagchi:2010}
A.~Bagchi,
  \href{https://link.aps.org/doi/10.1103/PhysRevLett.105.171601}{Correspondence
  between asymptotically flat spacetimes and nonrelativistic conformal field
  theories}, Phys. Rev. Lett. 105 (2010) 171601.
\newblock \href {http://dx.doi.org/10.1103/PhysRevLett.105.171601}
  {\path{doi:10.1103/PhysRevLett.105.171601}}.
\newline\urlprefix\url{https://link.aps.org/doi/10.1103/PhysRevLett.105.171601}

\bibitem{Bagchi:2013toa}
A.~Bagchi, R.~Gopakumar, {Non-relativistic AdS/CFT and the GCA}, J. Phys. Conf.
  Ser. 462~(1) (2013) 012017.
\newblock \href {http://dx.doi.org/10.1088/1742-6596/462/1/012017}
  {\path{doi:10.1088/1742-6596/462/1/012017}}.

\end{thebibliography}

\end{document}